\documentclass[twocolumn,english,aps,pra]{revtex4}
\usepackage[latin1]{inputenc}
\usepackage{graphicx}
\usepackage{amssymb}

\makeatletter

\usepackage{graphicx}

\usepackage{graphicx}

\usepackage{babel}
\makeatother

\begin{document}
\title{Semiclassical Wigner distribution for two-mode entangled 
state\\
generated by an optical parametric oscillator}
\author{K. Dechoum$^{1}$,  M. D. Hahn$^{1}$, 
        R. O. Vallejos$^{2}$, A. Z. Khoury$^{1}$ }
\affiliation{ ${}^{1}$ Instituto de F\'{\i}sica, Universidade Federal Fluminense,\\ 
Av. Gal. Milton Tavares de Souza s/n, 24210-346 Niter\'oi - RJ, Brazil}
\affiliation{ ${}^{2}$ Centro Brasileiro de Pesquisas F\'{\i}sicas, 
Rua Dr. Xavier Sigaud 150, 22290-180 Rio de Janeiro - RJ, Brazil}
\date{\today}

\begin{abstract}

We derive the steady state solution of the Fokker-Planck equation 
that describes the dynamics of the nondegenerate optical parametric
oscillator in the truncated Wigner representation of the density 
operator. 
We assume that the pump mode is strongly damped, which permits
its adiabatic elimination.
When the elimination is correctly executed, the resulting 
stochastic equations contain multiplicative noise terms,
and do not admit a potential solution.
However, we develop an heuristic scheme leading to a
satisfactory steady-state solution. 
This provides a clear view of the intracavity 
two-mode entangled state valid in all operating regimes of the 
OPO. 
A nongaussian distribution is obtained for the above threshold 
solution. 
\end{abstract} \pacs{} \maketitle

\section{Introduction}

Among the quasiprobability functions that represent the density 
operator of a quantum state, the Wigner distribution has undoubtedly 
advantages over others, since in this phase space representation the 
quantum-classical correspondence is, in general, much more visible. 
At the same time these functions contain all information available in 
the density operator.  

Unfortunately, just a few states have been represented by this 
distribution due to the difficulty to solve Fokker-Planck equations 
for nonlinear systems. 
Indeed, the phase space description of quantum systems is well 
known for quadratic hamiltonians \cite{carmichael,dodonov}, 
but very little is known for nonlinear 
systems, such as, for example, the single-mode degenerate parametric oscillator 
\cite{kinsler}, or the transverse multimode degenerate parametric oscillator 
\cite{kaled2}. 
 
Of special interest is the two-mode entangled state generated in 
the optical parametric oscillator (OPO) \cite{villar2}, used in many experiments 
of quantum information and recently shown to produce tripartite 
continuous variable entanglement \cite{science-usp}. 
The Wigner distribution of this state was obtained exactly using 
the solution derived from the complex P-representation 
(after adiabatic elimination of the pump mode) \cite{karen}, 
and tranformed into a 
Wigner distribution. 
Since this analytical result appears as an infinite series of gamma functions 
(hypergeometric series), some important physical aspects are hidden by its 
mathematical complexity. For example, two-mode entanglement is not of easy 
identification from the usual partial transpose criterion \cite{simon,duan}.
The same happens to the generalized P-representation for the intracavity 
parametric oscillator derived in Ref.~\cite{mcneil}. In order to calculate 
the intensity correlations, the authors obtained an integral expression 
that depends on degenerate hypergeometric functions, and the outcome also 
depends on an infinite sum. 

In the present work, we derive the Wigner function of a two-mode 
quantum state of the electromagnetic field generated by an OPO 
using a truncated Wigner equation. 
This truncated approximation is valid when the nonlinearity 
is small enough, which is true for most of the OPO experiments 
developed so far. 
A simple expression is obtained for the Wigner distribution, which 
renders clear the many quantum features of the intracavity OPO field 
such as squeezing and entanglement.
Moreover, this expression is valid in all regions of the OPO operation regimes, 
that is, below, and above threshold, where the validity 
of both the linear approximation, and the perturbation theory 
break down. 
This distribution also describes the OPO quantum fluctuations 
properly around the operation threshold, while the usual linear 
approach provides divergent results. 
In order to illustrate the reliability of the distribution we obtain, 
we compare moments calculated with this Wigner function with those obtained 
by numerical simulations using the positive P-representation 
of the density operator without any approximations, that is, full 
quantum result for any control parameters of the OPO.

The paper is organized as follows.
Section~\ref{sec2} presents the master equation describing the
three-mode OPO dynamics in the Wigner representation.
The correct adiabatic elimination of the pump mode is addressed
in Sec.~\ref{sec3}, where we show that the standard 
procedure \cite{graham} leads to wrong results for 
some quadrature moments.
After truncation and adiabatic elimination a two-mode Fokker-Planck
is obtained which does not satisfy the conditions for a 
steady-state potential solution. 
However, we show that an approximately potential solution 
can be found which reproduces satisfactorily many quantum
features of the OPO (Sec.~\ref{sec4}).
Comparisons with the linear theory are exhibited in 
Sec.~\ref{sec5}.
Section~\ref{sec6} contains the concluding remarks.

\section{Theoretical model}
\label{sec2}

We present here a model of three quantized modes coupled by a 
nonlinear crystal 
inside a triply resonant Fabry-Perot cavity.
The Heisenberg picture Hamiltonian that describes this open 
system is
given by \cite{carmichael}
\begin{eqnarray}
\hat{H} &=& \sum_{i=0}^{2} \hbar \omega_{i} \hat{a}_{i}^{\dagger} \hat{a}_{i}
+ i \hbar \chi \left(\hat{a}_{1}^{\dagger} \hat{a}_{2}^{\dagger}
\hat{a}_{0} - \hat{a}_{1} \hat{a}_{2} \hat{a}_{0}^{\dagger} \right) \nonumber \\
&+&  i\hbar \left( E e^{-i\omega_{0} t}\hat{a}_{0}^{\dagger} -
E^{*} e^{i\omega_{0} t}\hat{a}_{0} \right) \nonumber
\\
&+&\sum_{i=0}^{2} \left(
\hat{a}_{i} \hat{\Gamma}_{i}^{\dagger} + \hat{a}_{i}^{\dagger}
\hat{\Gamma}_{i} \right)\;.
\label{1}
\end{eqnarray}
Here $E$ represents the external coherent driving pump field at frequency 
$\omega_{0}$. The operators $\hat{a}_{0}$, $\hat{a}_{1}$ and $\hat{a}_{2}$ 
represent the pump, signal and idler fields, respectively, satisfying the 
following frequency matching condition, $\omega_{0}=\omega_{1}+\omega_{2}$. 
The terms $\hat{\Gamma}_{i}$ represent damping reservoir operators, 
and $\chi$ is the nonlinear coupling constant due to the second order 
polarizability of the nonlinear crystal.

The master equation for the reduced density operator, after the elimination 
of the heat bath by standard techniques \cite{carmichael}, is given by
\begin{eqnarray}
\frac{\partial \hat{\rho}}{\partial t} &=& 
-i \sum_{i=0}^{2} \omega_{i} \left[ \hat{a}_{i}^{\dagger}\hat{a}_{i}, \hat{\rho} \right] +
 \chi\left[ \hat{a}_{1}^{\dagger} \hat{a}_{2}^{\dagger}\hat{a}_{0}, 
\hat{\rho} \right] - 
 \chi\left[ \hat{a}_{1} \hat{a}_{2} \hat{a}_{0}^{\dagger},\hat{\rho} \right] \nonumber \\
&&+ E e^{-i\omega_{0} t}\left[ \hat{a}_{0}^{\dagger},\hat{\rho} \right]- 
 E^{*} e^{i\omega_{0} t}\left[ \hat{a}_{0},\hat{\rho} \right]\nonumber \\
&&+\sum_{i=0}^{2} \gamma_{i} \left( 2 \hat{a}_{i} \hat{\rho} 
\hat{a}_{i}^{\dagger} - \hat{a}_{i}^{\dagger} \hat{a}_{i} \hat{\rho} -
\hat{\rho} \hat{a}_{i}^{\dagger} \hat{a}_{i} \right)\;,
\label{2}
\end{eqnarray}

\noindent where $\gamma_{i}$ is the corresponding mode damping rate. 

In order to treat the operators evolution, we now turn to the method of 
operator representation theory. These techniques can be used to transform
the density matrix equation of motion into c-number Fokker-Planck or 
stochastic equations.
The phase space Wigner equation for the nondegenerate parametric amplifier 
that corresponds to the master equation (\ref{2}) is then 
\begin{eqnarray}
\frac{\partial W}{\partial t}&=& 
\left\{ 
\frac{\partial}{\partial \alpha_{0}} \left( i \omega_{0} \alpha_{0}+ 
\gamma_{0} \alpha_{0} + \chi \alpha_{1} \alpha_{2} - E e^{-i\omega_{0}t} 
\right)\right. \nonumber \\
&+& \frac{\partial}{\partial \alpha_{0}^{*}} 
\left(-i \omega_{0} \alpha_{0}^{*} + \gamma_{0} \alpha_{0}^{*}+
\chi \alpha_{1}^{*} \alpha_{2}^{*} - E^{*} e^{i\omega_{0}t}
\right)  \nonumber \\  
&+& \frac{\partial}{\partial \alpha_{1}} 
\left( i \omega_{1} \alpha_{1}+
\gamma_{1} \alpha_{1} - \chi \alpha_{2}^{*} \alpha_{0} \right) \nonumber \\
&+& \frac{\partial}{\partial \alpha_{1}^{*}} 
\left(-i \omega_{1} \alpha_{1}^{*} + 
\gamma_{1} \alpha_{1}^{*} - \chi \alpha_{2} 
\alpha_{0}^{*} \right)  \nonumber \\
&+& \frac{\partial}{\partial \alpha_{2}} 
\left( i \omega_{2} \alpha_{2}+
\gamma_{2} \alpha_{2} - \chi \alpha_{1}^{*} \alpha_{0} \right) \nonumber \\
&+& \frac{\partial}{\partial \alpha_{2}^{*}} 
\left(-i \omega_{2} \alpha_{2}^{*} +  
\gamma_{2} \alpha_{2}^{*} - \chi \alpha_{1} 
\alpha_{0}^{*} \right)  \nonumber \\
&+& \gamma_{0} \frac 
{\partial^{2}}{\partial \alpha_{0} \partial \alpha_{0}^{*}} +  
\gamma_{1} \frac 
{\partial^{2}}{\partial \alpha_{1} \partial \alpha_{1}^{*}} +  
\gamma_{2} \frac 
{\partial^{2}}{\partial \alpha_{2} \partial \alpha_{2}^{*}}  \nonumber \\ 
&+& \left. \frac{\chi}{4} \left(
\frac{\partial^{3}}{\partial \alpha_{1} \partial \alpha_{2} 
\partial \alpha_{0}^{*}} +  
\frac{\partial^{3}}{\partial \alpha_{1}^{*}\partial 
\alpha_{2}^{*} \partial \alpha_{0}} \right)
\right\} W\;.
\label{5}
\end{eqnarray}  

This is not a Fokker-Planck equation due to the third order 
derivative term, but in the case where $\chi$ is small enough 
we can drop this term, as discussed in appendix A 
(see Ref.~\cite{polvo} for a recent review of the truncated 
Wigner approximation and its applications).
The truncated equation so obtained is a genuine Fokker-Planck 
equation with a positive diffusion term, 
and we can easily derive the corresponding stochastic differential 
equations in the rotating frame 
($\tilde{\alpha}_j = \alpha_j\exp{(-i\omega_j t)}$ with $j=0,1,2$):
\begin{eqnarray}
&& \frac{d \tilde{\alpha}_{0}}{dt} = -\gamma_{0} \tilde{\alpha}_{0} + E - 
\chi \tilde{\alpha}_{1} \tilde{\alpha}_{2} + \sqrt{\gamma_{0}} \xi_{0}(t)
\;,\nonumber \\ 
&& \frac{d \tilde{\alpha}_{1}}{dt} = -\gamma_{1} \tilde{\alpha}_{0} + 
\chi \tilde{\alpha}_{0} \tilde{\alpha}_{2}^{*} + \sqrt{\gamma_{1}} \xi_{1}(t) 
\;,\nonumber \\
&& \frac{d \tilde{\alpha}_{2}}{dt} = -\gamma_{2} \tilde{\alpha}_{0} + 
\chi \tilde{\alpha}_{0} \tilde{\alpha}_{1}^{*} + \sqrt{\gamma_{2}} \xi_{2}(t) \;.
\end{eqnarray}

\section{Validity of the adiabatic approximation}
\label{sec3}

It is possible to find a stationary solution of the truncated 
Fokker-Planck equation when we adiabatically eliminate the pump 
mode variables. 
This means that we are considering the relaxation rate of this 
mode much larger than those of the downconverted modes, that is 
$\gamma_{0} \gg \gamma_{1},\gamma_{2}$. 
The stationary solution for the pumped mode is 
\begin{equation}
\tilde{\alpha}_{0}= 
\frac{1}{\gamma_{0}} \left[ E - \chi \tilde{\alpha}_{1} 
\tilde{\alpha}_{2} + \sqrt{\gamma_{0}} \xi_{0} (t) \right]\;,
\label{8}
\end{equation}
where the noise term is retained in the adiabatic elimination 
in order to properly deal with the noise dynamics. 
We are then left with two nonlinear dynamical equations
for the complex amplitudes of the down-converted fields, 
where the pump amplitude 
is replaced by expression (\ref{8}).

We now define the following real quadrature variables:
\begin{eqnarray}
x_{1} =  \tilde{\alpha}_{1} + \tilde{\alpha}_{1}^{*}\;, \,\,\,\,\,\,\,\,\,\, 
y_{1} = -i\left( \tilde{\alpha}_{1} - \tilde{\alpha}_{1}^{*} \right )\;,
\nonumber \\
x_{2} =  \tilde{\alpha}_{2} + \tilde{\alpha}_{2}^{*}\;, \,\,\,\,\,\,\,\,\,\, 
y_{2} = -i\left( \tilde{\alpha}_{2} - \tilde{\alpha}_{2}^{*} \right )\;,
\end{eqnarray}
so that the remaining dynamical equations can be cast 
in the compact form:
\begin{equation}
\frac{d \mathbf{X}}{dt} = \mathbf{A} + \mathbf{B}\,\mathbf{\xi}(t)\;,
\label{langevin}
\end{equation}
where $\mathbf{X}=[x_1,y_1,x_2,y_2]^T$ is a column vector, 
and the drift vector is defined as 
\begin{equation}
\mathbf{A} = \gamma \left[
\begin{array}{c}
-x_{1} + \mu x_{2} - 
\frac{g^{2}}{2} x_{1} \left(x_{2}^{2} +y_{2}^{2}\right)  \\
\noalign{\medskip}
-y_{1} - \mu y_{2} - 
\frac{g^{2}}{2} y_{1} \left(x_{2}^{2} +y_{2}^{2}\right)  \\
\noalign{\medskip}
-x_{2} + \mu x_{1} - 
\frac{g^{2}}{2} x_{2} \left(x_{1}^{2} +y_{1}^{2}\right)  \\
-y_{2} - \mu y_{1} - 
\frac{g^{2}}{2} y_{2} \left(x_{1}^{2} +y_{1}^{2}\right)  \\
\end{array}
\right]\;.
\end{equation}
We have set $\gamma_{1}=\gamma_{2}=\gamma$, 
which is a reasonable physical assumption 
for most OPO experiments, and defined the normalized 
pump parameter $\mu = \chi E/(\gamma \gamma_{0})$, 
as well as the nonlinear coupling 
$g = \chi/(\sqrt{2 \gamma \gamma_{0}})$.  
$\mathbf{B}$ is a $4\times 6$ matrix defined as
\begin{equation}
\mathbf{B} = \sqrt{2 \gamma} \left(
\begin{array}{cccccc}
1 &0 &0 &0 & \frac{g}{\sqrt{2}}x_{2} & \frac{g}{\sqrt{2}}y_{2} \\
0 &1 &0 &0 & -\frac{g}{\sqrt{2}}y_{2} & \frac{g}{\sqrt{2}}x_{2} \\
0 &0 &1 &0 & \frac{g}{\sqrt{2}}x_{1} & \frac{g}{\sqrt{2}}y_{1} \\
0 &0 &0 &1 & -\frac{g}{\sqrt{2}}y_{1} & \frac{g}{\sqrt{2}}x_{1} \\
\end{array}
\right)
\end{equation}
and $ {\mathbf \xi}(t) $ is a six component column vector 
whose entries are uncorrelated real 
Gaussian noises associated with the noise terms for 
the three interacting fields. 

Note that the last two columns in $\mathbf{B}$ give rise to 
multiplicative noise terms in the 
stochastic equation (\ref{langevin}). 
It is important to remark that 
if these multiplicative noise terms are drop \cite{graham}, 
some essential quantum features, particularly important in 
the above threshold regime, will be lost. 

\begin{figure}[htp]
\includegraphics[scale=0.3]{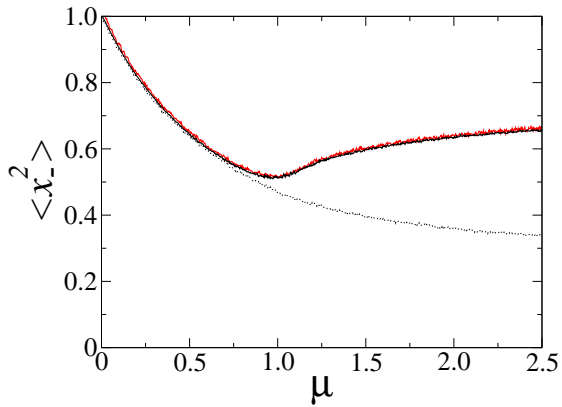}
\caption{Squeezed quadrature variance 
as a function of the pump level. 
We compare the results of the positive-P (solid black line) and 
truncated-Wigner stochastic equations with (solid red line) 
and without (dotted black line) multiplicative noise terms.
The first two curves are almost indistinguishable; 
the third one decreases monotonically. We chose $g^2=0.01$ in all 
calculations and $\gamma_0=10\,\gamma$ in the positive-P simulation. 
All plotted quantities are dimensionless.} 
\label{comparison}
\end{figure}

In order to illustrate the importance of the multiplicative 
noise terms around and above threshold, and study 
the validity of both the truncation and adiabatic 
approximations, we compare the numerical 
integration of the stochastic equations in the 
positive-P representation (without adiabatic elimination) 
with the stochastic equations 
for the truncated Wigner representation 
in the adiabatic approximation both with and without 
multiplicative noise terms. 
The results for the squeezed quadrature variances 
(defined in Sec.~\ref{sec5}) as a 
function of the pump level is presented in 
Fig.~\ref{comparison}. 
The positive-P calculation clearly agrees with the truncated 
Wigner if the multiplicative noise terms are properly taken 
into account.
Although valid below threshold, neglection of the multiplicative 
noise terms clearly fails around and above threshold, where it 
predicts the unphysical \cite{wallsmilburn} vanishing of the 
intracavity quadrature noise. 

\section{Approximate potential solution of the Fokker-Planck Equation}
\label{sec4}

The set of stochastic equations (\ref{langevin}) can be mapped onto a genuine Fokker-Planck equation:
\begin{equation}
\frac{\partial W(\mathbf{X})}{\partial t} = \left[ -\frac{\partial}{\partial \mathbf{X}_{i}} \mathbf{A}_{i} + 
\frac{1}{2}\frac{\partial}{\partial \mathbf{X}_{i}} \frac{\partial}{\partial \mathbf{X}_{j}} \mathbf{D}_{ij}\right]
W(\mathbf{X})
\label{10}
\end{equation}
where the diffusion matrix is defined as $\mathbf{D}= \mathbf {B} \mathbf {B}^{T}$, and summation over repeated 
indices is assumed throughout the text \cite{carmichael}.

Neglection of the multiplicative noise terms in the adiabatic elimination allows for 
a steady state solution in the 
potential form $W(\mathbf{X})= N \exp{(-\int \mathbf{Z}_{i}\,d\mathbf{X}_{i})}$ with $\mathbf{Z}_{i}$ 
given by
\begin{equation}
\mathbf{Z}_{i} = \mathbf{D}_{ik}^{-1} \left[2 \mathbf{A}_{k} - \frac{\partial}{\partial \mathbf{X}_{j}}
\mathbf{D}_{kj} \right]\;,
\end{equation}
as obtained in Ref.~\cite{graham}. 
As we have seen, this distribution gives incorrect squeezing 
results for above threshold operation. 
On the other side, if multiplicative noise terms are kept,
a potential solution is no longer possible 

This difficulty can be circumvented if one replaces the 
diffusion matrix by its mean value, 
as explained in Appendix B, so that a potential solution 
for $W(\mathbf{X})$ becomes possible.
This solution provides a quite simple form for the Wigner distribution which allows for the calculation of 
several statistical properties. 
We remark that this approximate potential solution is achievable 
only when $\gamma_1=\gamma_2\equiv\gamma$ as we have already 
assumed \cite{gardiner}. 
In this case, the steady state Wigner distribution for signal and idler reads:
\begin{eqnarray}
&&W(x_{1},y_{1},x_{2},y_{2}) = {\cal{N}} \exp \left\{\frac{-1}{2 s(\mu)} 
\left[ x_{1}^{2}+ y_{1}^{2}+x_{2}^{2}+ y_{2}^{2} \right.\right.
\nonumber\\
&&\left.\left. + 2 \mu \left( y_{1}y_{2} -x_{1}x_{2} \right)
 + g^{2} \left(x_{1}^{2} +y_{1}^{2}\right) 
\left(x_{2}^{2} +y_{2}^{2}\right) \right]\right\} 
\label{11}
\end{eqnarray}
where ${\cal{N}}$ is the normalization constant and unimportant 
terms $O(g^4)$ were neglected.
We also defined 
\begin{equation}
s(\mu) = \left\{
\begin{array}{cc}
1   & (\mu\leq 1) \\
\mu & (\mu\geq 1) \\
\end{array}
\right.\;.
\label{eq15}
\end{equation}

As a concrete physical example, expression (\ref{11}) can describe the 
joint Wigner distribution for two polarization modes of a frequency 
degenerate type II OPO, where signal and idler are distinguished by 
their polarization states. 
All statistical properties, including experimentally accessible quantities 
like quadrature noise and correlations in the stationary state may be 
calculated with this distribution in any operation regime of the OPO 
within the validity of the adiabatic elimination. 
It is interesting to 
note that the distribution given by 
expression (\ref{11}) is single peaked below threshold 
operation ($\mu < 1$) and becomes double peaked above threshold. 
In Fig.~2, it is possible to visualize the conditional Wigner 
distribution for fixed values of $y_{1}$ and $y_{2}$, for an 
OPO operating below, at, and above threshold. It is easy to identify 
the squeezed and unsqueezed quadratures in all regimes.

\begin{figure}
\resizebox*{8.0cm}{5cm}{\rotatebox{0}{\includegraphics{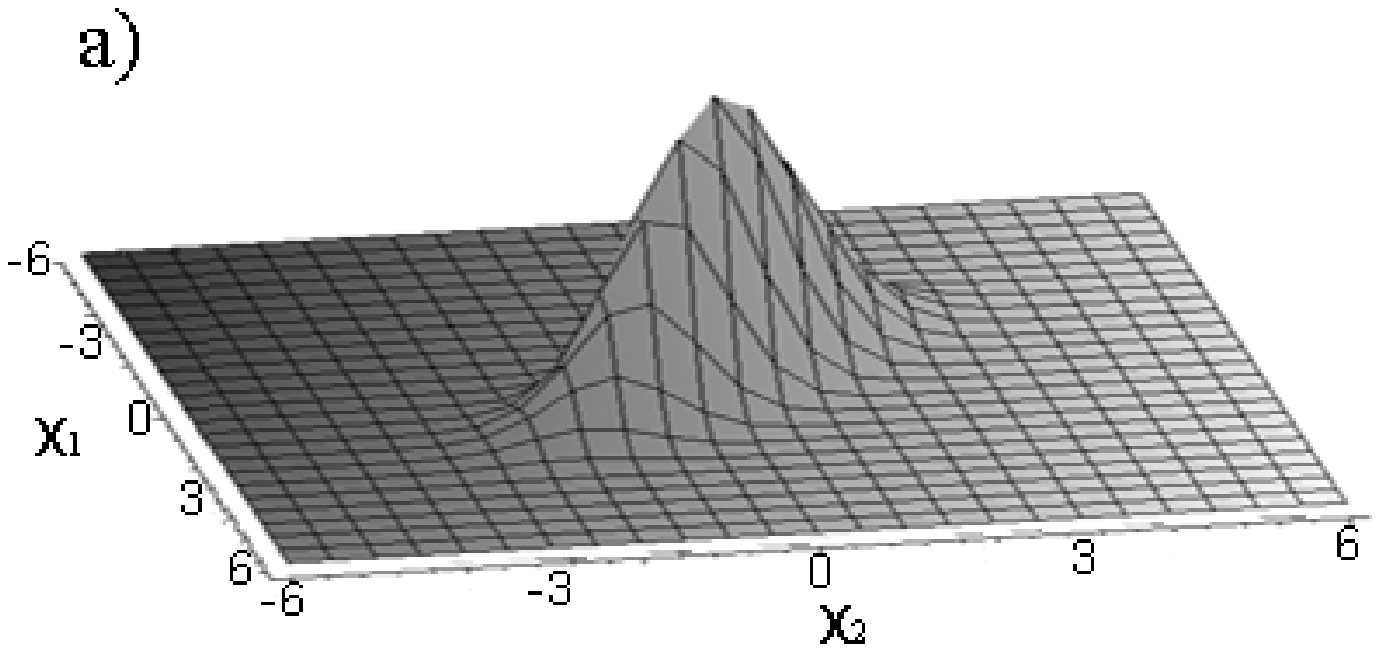}}}\\
\resizebox*{8.0cm}{5cm}{\rotatebox{0}{\includegraphics{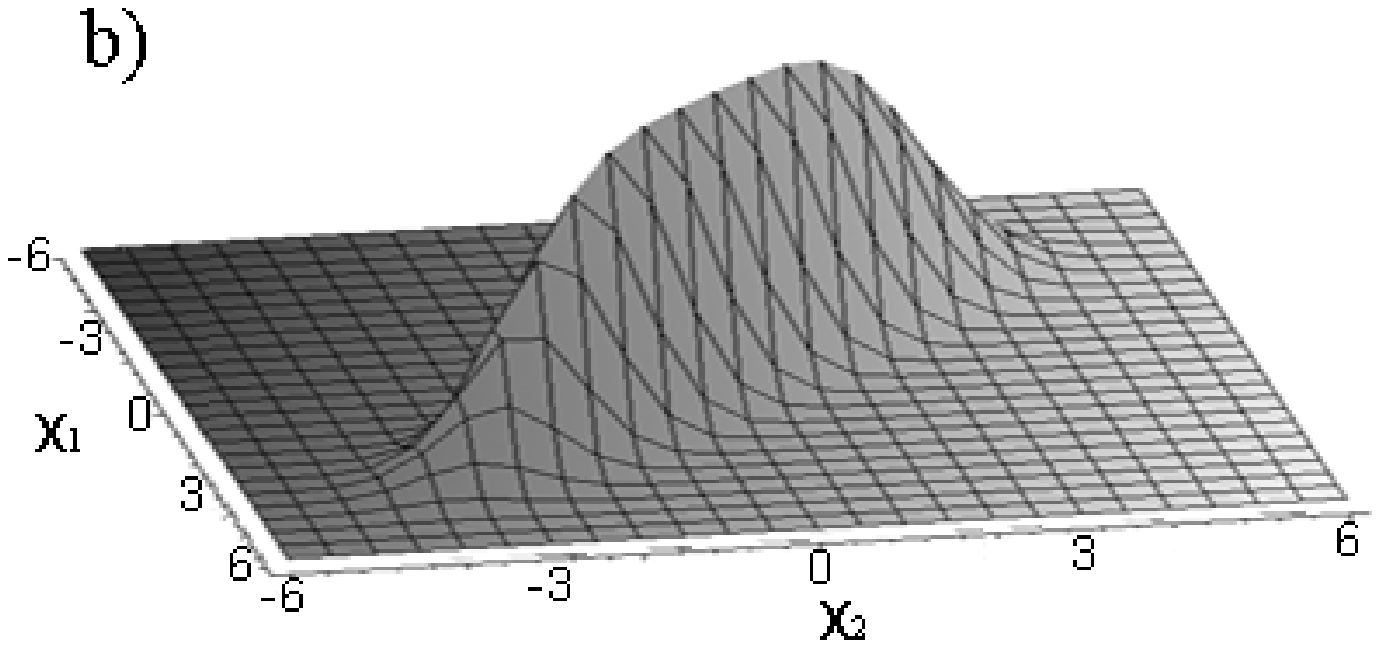}}}\\
\resizebox*{8.0cm}{5cm}{\rotatebox{0}{\includegraphics{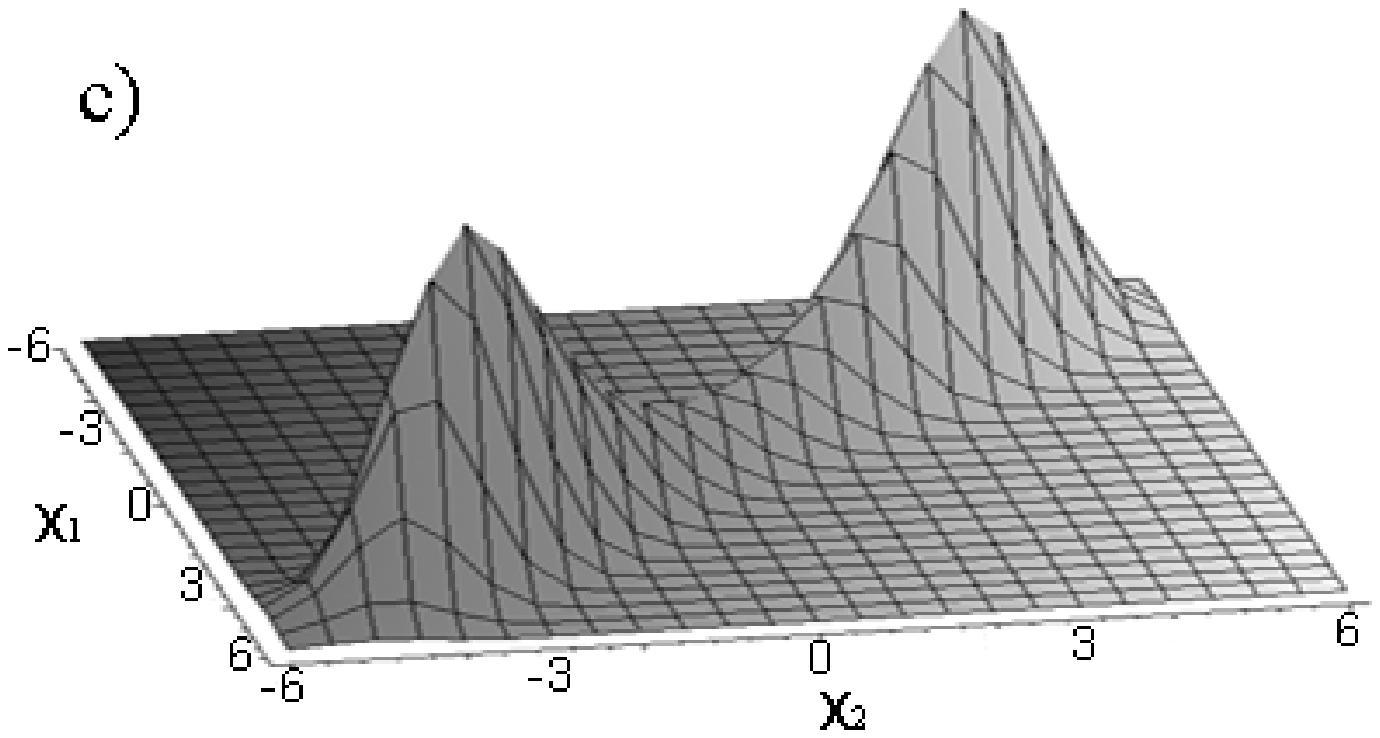}}}
\caption{Conditional Wigner distribution for $y_{1}=0$ and $y_{2}=0$, 
a) below ($\mu = 0.5$); b) at ($\mu = 1.0$); and c) above threshold 
($\mu = 1.5$). In all cases $g^{2}=0.01$. All plotted quantities are 
dimensionless.}
\label{fig2}
\end{figure}

For the nonlinear media employed in most OPO devices, the coupling 
parameter $g\ll 1$. Therefore, the first two terms in the 
exponent of the Wigner distribution governs the OPO behaviour 
below threshold. They are enough to predict the usual 
nonclassical features. In fact, it is evident now that the usual 
linearized approach to quantum noise 
in below-threshold OPOs is equivalent to completely neglect the 
$g^2$ term in the distribution. 
However, it is important to notice 
that for $\mu\geq 1$, the Wigner distribution becomes divergent if 
we neglect the $g^2$ term. This means that this term plays a 
crucial role for a complete description of the OPO behaviour 
at and above threshold. 

\begin{figure}[ht]
\resizebox*{8.0cm}{5cm}{\rotatebox{0}{\includegraphics{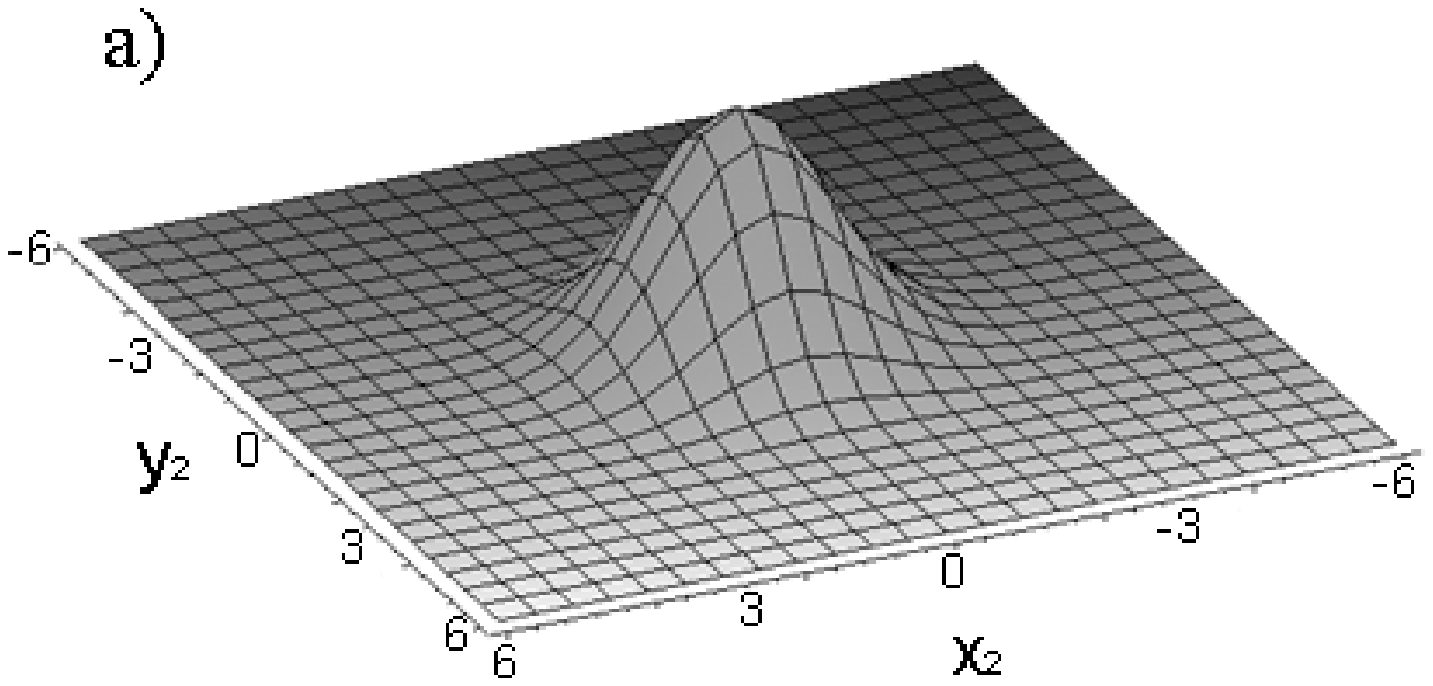}}}
\resizebox*{8.0cm}{5cm}{\rotatebox{0}{\includegraphics{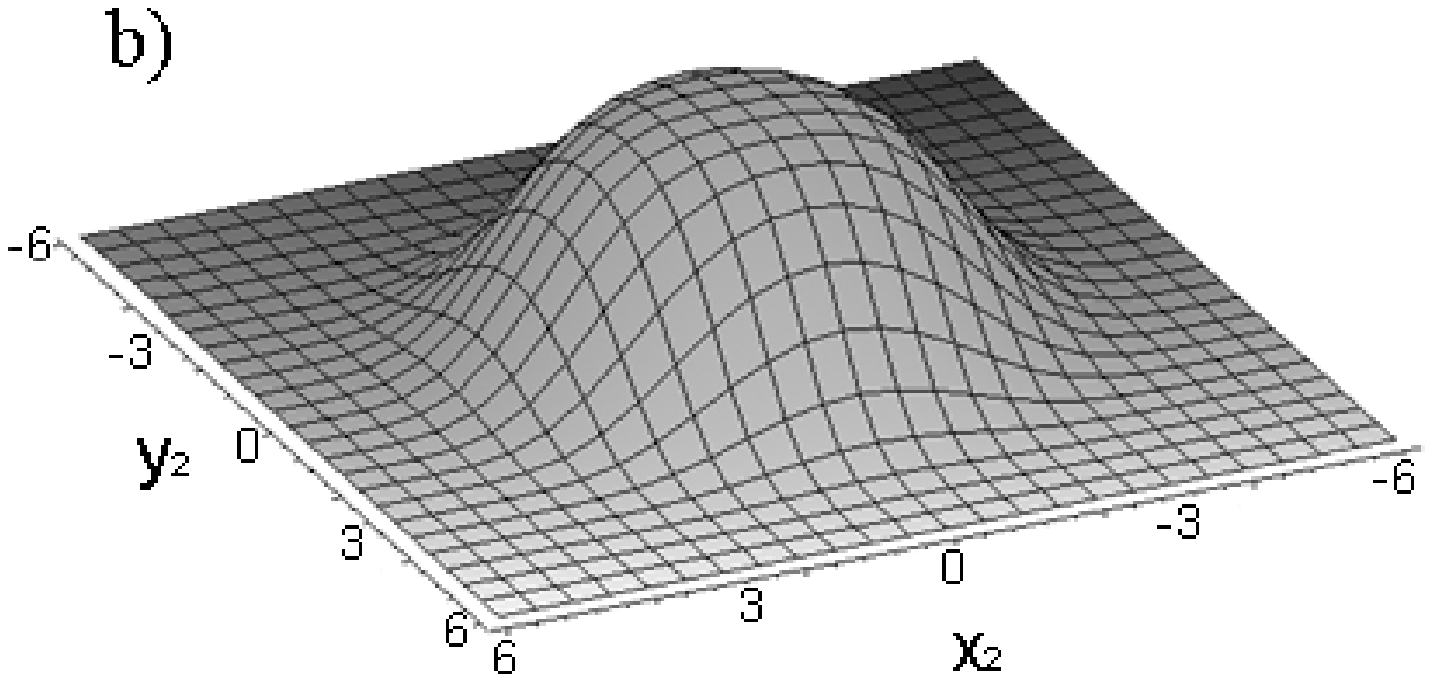}}}
\resizebox*{8.0cm}{5cm}{\rotatebox{0}{\includegraphics{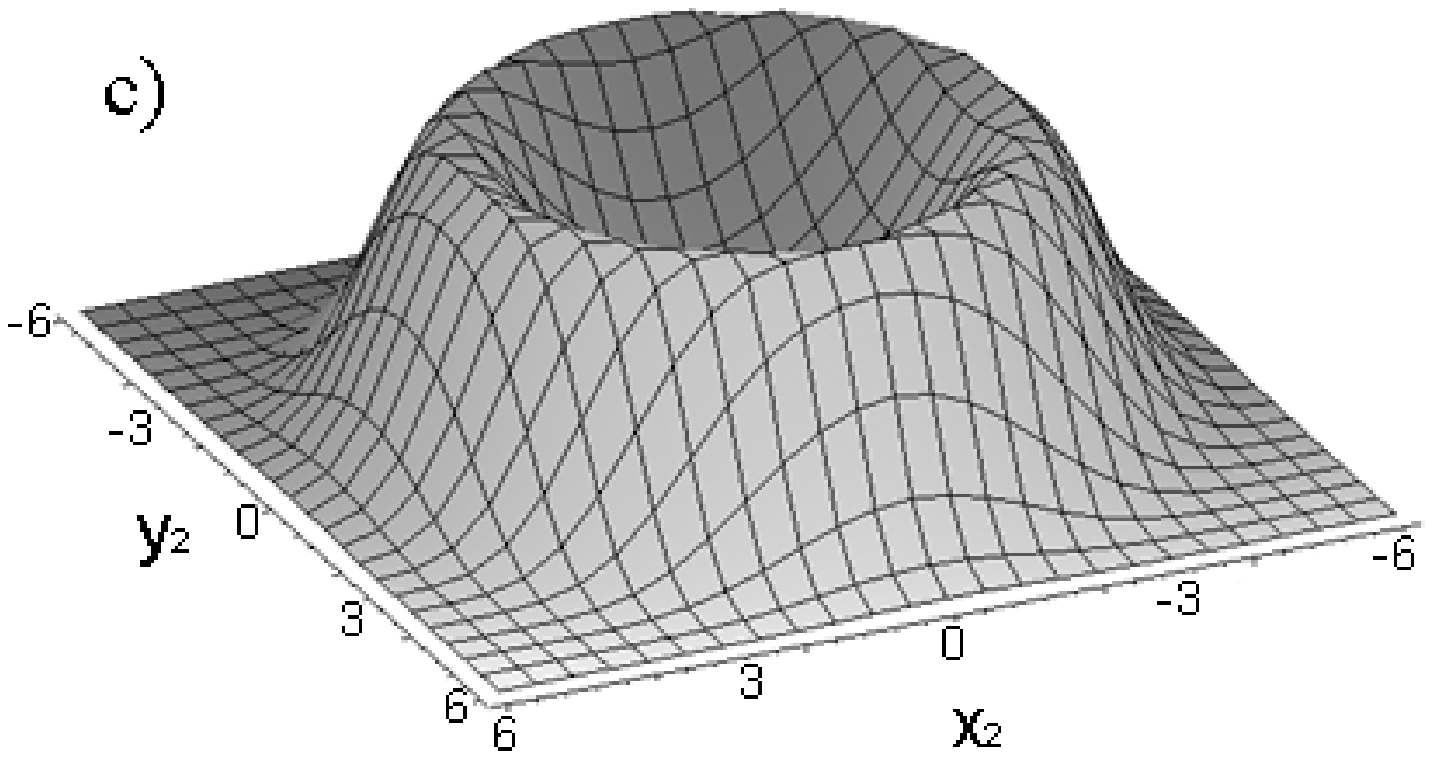}}}
\caption{Marginal distributions: a) below threshold, $\mu = 0.8$; 
b) at threshold, $\mu = 1.0$; 
c) above threshold, $\mu = 1.2$. In all cases $g^{2}=0.01$. 
All plotted quantities are dimensionless.}
\label{fig3}
\end{figure}

Additional insight is provided by a careful investigation of the marginal 
distribution obtained by integrating $W(x_{1}, y_{1},x_{2}, y_{2})$ with 
respect to one mode variables. Of course, due to the symmetry of $W$, the 
form of the resulting marginal distribution is independent of the mode 
being traced out. The marginal distribution for mode 2 becomes:
\begin{eqnarray}
&&W\left(x_2,y_2\right)=\frac{2\pi {\cal{N}}\mu}{1+g^2\left(x_{2}^{2}+y_{2}^{2}\right)}\times
\label{marginal}\\ 
&&\exp\left\{\frac{-1}{2s(\mu)}\left[\left(x_{2}^2+y_{2}^{2}\right)
\left(1-\mu^2\right)+g^2\left(x_{2}^2+y_{2}^{2}\right)^2\right]\right\}
\nonumber
\end{eqnarray}
It gives the statistical properties of isolated measurements 
on mode 2.
The variances given by this marginal distribution are larger 
than those of the vacuum state. 
This excess noise can be seen as a consequence of 
information loss when one looks only at part of the whole 
system.
Fig.~\ref{fig3} presents the marginal distribution below, 
at and above threshold. 
Above threshold the marginal distribution is clearly peaked 
out of the 
phase space origin, what is a consequence of the 
macroscopic amplification 
of the down-converted fields. 
However, the distribution presents radial 
symmetry indicating complete phase uncertainty in each 
individual mode.

\section{Comparison with the linearized quantum approach} 
\label{sec5}

Almost the totality of theoretical works devoted to the analysis of 
quantum noise in optical systems rely on linearization of the small 
quantum fluctuations around the macroscopic steady state mean values. 
It is worth to notice that this procedure has limited validity, especially 
around the threshold critical point where quantum fluctuations may 
become comparable to the mean values. In order to evidence the 
breakdown of the linearized theory, we shall compare its results with 
those obtained from the solution of the Fokker-Planck equation (\ref{10}). 

The usual approach given to quantum noise in the literature is obtained 
from first order stochastic equations of motion \cite{ndturco,villar}.
These equations are often used to predict squeezing in a linearized 
fluctuation analysis. They are non-classical in the sense that they 
can describe states without a positive Glauber-Sudarshan P-distribution, 
but correspond to a Gaussian Wigner distribution. 

We now find it useful to introduce combined field quadratures, as in 
two-mode approaches used previously~\cite{Caves}. These combined quadratures 
are the Einstein-Podolsky-Rosen (EPR) variables used to characterize continuous 
variable entanglement between the down converted fields.  
They are defined as
\begin{equation}
x_{\pm} = \frac{x_{1} \pm x_{2}} {\sqrt{2}}, \;\;\;\;\;\; 
y_{\pm} = \frac{y_{1} \pm y_{2}} {\sqrt{2}}.
\label{eq:newquads}
\end{equation}
These quantities correspond to the squeezed and anti-squeezed combined 
quadratures obtained in the linearized theory. 
From the first order stochastic equations one can easily obtain the 
steady state variances of the EPR variables \cite{ndturco,villar}:
\begin{eqnarray}
\langle x_{+}^{2}\rangle &=& \langle y_{-}^{2}\rangle = \frac{1}{1-\mu}
\nonumber\\
\langle x_{-}^{2}\rangle &=& \langle y_{+}^{2}\rangle = \frac{1}{1+\mu}
\label{eq:variancesbelow}
\end{eqnarray}
for below threshold operation, and
\begin{eqnarray}
\langle x_{+}^{2}\rangle &=& \langle y_{-}^{2}\rangle = \frac{1}{\mu-1}
+ \frac{1}{g^2}(\mu-1)
\nonumber\\
\langle x_{-}^{2}\rangle &=& \langle y_{+}^{2}\rangle = \frac{1}{2}
\label{eq:variancesabove}
\end{eqnarray}
for above threshold operation. 

Now let us briefly discuss the predictions of the linearized approach 
and its validity. The quadratures $x_-$ and $y_+$ exhibit 
the expected squeezing, going from the vacuum fluctuations for zero 
pump until 50\% intracavity squeezing (which corresponds to perfect 
squeezing outside the cavity) at and above the oscillation threshold.
The unsqueezed quadratures $x_+$ and $y_-$ present divergent behaviour 
around threshold, which is certainly not physical. In fact, as we 
shall see next, the steady state solution of the Fokker-Planck equation 
(\ref{10}) gives a well behaved Wigner distribution which does not 
display any divergences. This Wigner distribution will also put limits 
on the amount of squeezing attainable.

In order to investigate the two-mode entanglement directly from the Wigner 
distribution, we now write it in terms of the EPR variables.
Below threshold, where the linearization of the equation of motion is valid, 
we can neglect the $g^{2}$ term, and the distribution can be approximated as 

\begin{eqnarray}
&&W_L(x_{+},y_{+},x_{-},y_{-}) = {\cal{N}} 
exp \left\{-\frac{1}{2}\left[ \left(1+\mu \right) x_{-}^{2} +\right.\right.
\nonumber\\ 
&&\left.\left.\left(1+\mu \right)y_{+}^{2}+ \left(1-\mu \right) x_{+}^{2}+ \left(1-\mu \right) y_{-}^{2} 
\right] \right\}\;. 
\label{weprL}
\end{eqnarray}

With this expression we can easily calculate any moment 
of the distribution. 
For instance, we easily 
reobtain the results of Eqs.~(\ref{eq:variancesbelow}) 
for the intracavity noise squeezing in the combined 
quadratures.
As we mentioned before, at threshold we have perfect external squeezing while the unsqueezed combined 
quadratures blow up showing the failure of linearization \cite{ndturco}. 
Moreover, it is easy to characterize this state as 
entangled by using the Duan-Simon criterion \cite{simon,duan}. Above threshold we can also calculate the moments 
and we find that this criterion shows the sufficient condition to characterize this state as entangled, 
in spite of not being a Gaussian state in that regime.

\begin{figure}
\resizebox*{7.0cm}{5.3cm}{\includegraphics{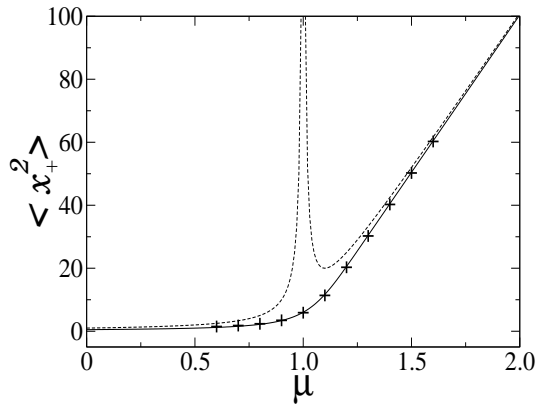}}
\caption{Comparison between the values of the anti-squeezed 
variable $\langle x^2_+\rangle$ obtained from the linearized 
theory (dashed line) with those obtained from the Wigner 
function of reference \cite{karen} (solid line), and the one 
given by Eq.~(\ref{11}) (crosses). In all cases we have set 
$g^2=0.01$. All plotted quantities are dimensionless.}
\label{limiar}
\end{figure}

\begin{figure}
\resizebox*{7.0cm}{5.3cm}{\includegraphics{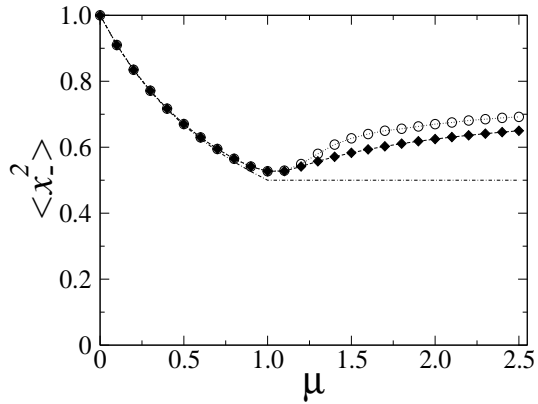}}
\caption{Comparison between the values of the squeezed 
variable $\langle x^2_-\rangle$ obtained from the linearized 
theory (solid line) with those obtained from the Wigner 
function of Ref.~\cite{karen} (open circles), 
and the one given by Eq.~(\ref{11}) (rhombs). 
In all cases we set $g^2=0.01$. 
All plotted quantities are dimensionless.}
\label{limiar2}
\end{figure}

It is also interesting to compare the results obtained 
for $\langle x^2_+\rangle$ 
with the linearized theory (both below and above threshold) 
with those obtained 
from the Wigner function given by Eq.~(\ref{11}). 
This is done in Fig.~(\ref{limiar}). 
While the linearized approach gives a divergent behavior at 
the oscillation 
threshold, the results obtained from the truncated Wigner 
function gives a more 
realistic smooth behavior in agreement with the full 
quantum solution of 
Refs.~\cite{karen,mcneil,singh}. 
The results for the squeezed quadrature $\langle x^2_-\rangle$ 
are shown in Fig.~\ref{limiar2} 
where, again, the results given by the truncated Wigner 
distribution agree with those obtained 
from the full quantum solution. Below threshold, the 
linearized approach also agrees with the 
full quantum theory and the truncated Wigner approach, 
but an important deviation can be 
observed above threshold where the linearized theory gives 
$\langle x^2_-\rangle=0.5$ for 
any pump power. 
The same results are obtained for the other squeezed variable 
$\langle y^2_+\rangle$, so that the Duan-Simon separability 
criterion \cite{simon,duan} shows 
that the two-mode quantum state is entangled. 
However, 
note that both 
the full quantum approach and the truncated Wigner distribution 
predict a less pronounced 
violation of the criterion above threshold.

In Refs.~\cite{mcneil,singh} Fokker-Planck equations were 
derived for the complex-P \cite{mcneil} and positive-P 
\cite{singh} representations, and stationary solutions were 
obtained for the corresponding distributions. 
Since these Fokker-Planck equations were derived without 
any truncation, we can consider the stationary distributions 
so obtained as \textit{exact}. 
In Ref.~\cite{karen} an exact stationary Wigner function was 
mapped from the complex-P distribution of Ref.~\cite{mcneil} 
and expressed in terms of Bessel functions. 
It is important to remark that the simple Wigner function 
presented here provides a very good approximation for those 
exact solutions, as can be seen in Figs.~\ref{limiar} and 
\ref{limiar2}, even for a rather large value for the 
nonlinear coupling $g^2$. 

\section{Conclusion}
\label{sec6}

By truncating the evolution equation for the 
Wigner function of an optical parametric oscillator, 
followed by the proper adiabatic elimination of the pump 
mode, we obtained a Fokker-Planck equation which describes 
satisfactorily the properties of the steady state 
in any operation regime of the OPO 
within the validity of the adiabatic elimination. 
This was checked by direct numerical simulation of the
corresponding stochastic equations.
Then we showed that the Fokker-Planck admits a simple
``quasi-potential" solution. 
For the unsqueezed quadratures, this solution 
does not present the unphysical divergent behavior of the 
linearized theory on the oscillation threshold. 
Moreover, it provides good agreement with exact 
quasiprobabilities already available in the literature, 
both for the squeezed and unsqueezed quadratures. 
While these exact solutions are given as infinite series, 
the potential solution of the truncated Wigner distribution 
presents a simple form allowing for an easier visualization 
of the phase space distribution.

\begin{acknowledgments}

We thank Cesar Raitz for useful comments.
This work was supported by the Instituto Nacional de 
Ci\^encia e Tecnologia de 
Informa\c c\~ao Qu\^antica (INCTIQ - CNPq - Brazil), 
Coordena\c c\~ao de 
Aperfei\c coamento de Pessoal de N\'{\i}vel Superior (CAPES - Brazil) and Funda\c c\~ao 
de Amparo \`a Pesquisa do Estado do Rio de Janeiro (FAPERJ).

\end{acknowledgments}

\section*{APPENDIX A} 
In order to clarify the hierarchy of the terms in the 
dynamical equation for the Wigner distribution, let us 
define the following dimensionless parameters:
\begin{itemize}
\item Dimensionless coupling:
\begin{equation}
g\equiv \frac{\chi}{\sqrt{2\gamma\gamma_0}}
\end{equation}
\item Dimensionless pump:
\begin{equation}
\mu\equiv \frac{E}{E_c}\equiv\frac{\chi E}{\gamma\gamma_0}
\end{equation}
\item Dimensionless time:
\begin{equation}
\tau\equiv \gamma\,t
\end{equation}
\item Dimensionless damping:
\begin{equation}
\gamma_r\equiv \frac{\gamma_0}{\gamma}
\end{equation}
\item Rescaled field amplitudes:
\begin{eqnarray}
\beta_0&\equiv& g\,\sqrt{2\gamma_r}\,\alpha_0\,e^{i\omega_0\,t} \\ 
\beta_1&\equiv& g\,\alpha_1\,e^{i\omega_1\,t} \\ 
\beta_2&\equiv& g\,\alpha_2\,e^{i\omega_2\,t}
\end{eqnarray}
\end{itemize}
with $\omega_0 = \omega_1 + \omega_2$ (phase match) and $E_c\equiv\gamma\gamma_0/\chi$ 
(threshold pump amplitude). 

Perturbative approaches to the quantum stochastic equations 
of the parametric oscillator 
has been widely discussed in Refs.~\cite{ndturco}. 
In these approaches, 
the perturbation parameter is the dimensionless coupling $g$, usually assumed 
to be small as is the case in most experimental situations. The quantum 
fluctuations are then expanded in powers of $g$ in order to provide progressive 
corrections to linearized theories (the zero order terms correspond to the 
macroscopic mean field equations). 

Therefore, 
let us rewrite the phase space Wigner equation for the 
nondegenerate parametric amplifier in terms of the rescaled amplitudes and 
dimensionless parameters. The equation then becomes:
\begin{eqnarray}
\frac{\partial W}{\partial \tau}&=& 
\left\{\gamma_{r}  
\frac{\partial}{\partial \beta_{0}} \left( 
\beta_{0} + 2\,\beta_{1} \beta_{2} - \mu 
\right)\right.\nonumber\\
&+& \gamma_{r} 
\frac{\partial}{\partial \beta_{0}^{*}} 
\left(\beta_{0}^{*}+
2\,\beta_{1}^{*} \beta_{2}^{*} - \mu
\right) 
\nonumber \\  
&+& \frac{\partial}{\partial \beta_{1}} 
\left( \beta_{1} - \beta_{2}^{*} \beta_{0} \right) 
+ \frac{\partial}{\partial \beta_{1}^{*}} 
\left(\beta_{1}^{*} - \beta_{2} 
\beta_{0}^{*} \right)\label{5b}\\ 
&+& \frac{\partial}{\partial \beta_{2}} 
\left( \beta_{2} - \beta_{1}^{*} \beta_{0} \right) 
+ \frac{\partial}{\partial \beta_{2}^{*}} 
\left(\beta_{2}^{*} - \beta_{1} 
\beta_{0}^{*} \right)
\nonumber \\
&+& g^2 \left( 2\,\gamma_{r}^{2} \frac 
{\partial^{2}}{\partial \beta_{0} \partial \beta_{0}^{*}} +  
\frac{\partial^{2}}{\partial \beta_{1} \partial \beta_{1}^{*}} +  
\frac{\partial^{2}}{\partial \beta_{2} \partial \beta_{2}^{*}}\right)\nonumber\\ 
&+& \left. g^4\frac{\gamma_r}{\sqrt{2}} \left(
\frac{\partial^{3}}{\partial \beta_{1} \partial \beta_{2} 
\partial \beta_{0}^{*}} +  
\frac{\partial^{3}}{\partial \beta_{1}^{*}\partial 
\beta_{2}^{*} \partial \beta_{0}} \right)
\right\} W\;.
\nonumber
\end{eqnarray}  

This rescaled form, evidences the hierarchy of the different 
terms in the dynamical Wigner equation and makes it clear that 
the third order derivative terms are negligible for small 
coupling, since it is proportional to $g^4$.
It is interesting to make a close inspection in the physical 
meaning of this hierarchy. 
For small enough coupling ($g\ll 1$), if we neglect both 
the $g^2$ and $g^4$ terms, we are left with a Liouvillian 
evolution which is deterministic in essence, any indeterminacy 
comes from the initial conditions. 
In this case, for example, initially delta distributed 
amplitudes evolve as such for all times:
\begin{equation}
W(t) = C\, \delta(\beta_0-\beta_0(t))\,\delta(\beta_1-\beta_1(t))\,\delta(\beta_2-\beta_2(t))\;.
\end{equation}
This corresponds to a classical deterministic evolution of the system. 

As $g$ is continuously increased, the next approximation 
corresponds to neglecting only the $g^4$ term. 
In this case, Eq.~(\ref{5b}) assumes the form 
of a standard Fokker-Planck equation, leading to the usual 
diffusive evolution 
with a possible stationary state (under self pulsing conditions, 
for instance, 
there is no stationary solution). 

Finally, there are effects of the $g^4$ term which are 
essentially quantum mechanical, but have received little 
attention in the literature. 
However, 
these effects require a coupling constant consideraby 
larger than what is 
experimentally available in parametric systems.

\section*{APPENDIX B} 
Here we clarify the steady state solution derivation of 
the effective Fokker-Planck equation for the two 
down-converted modes of the field, after adiabatic elimination 
of the pump. 
We begin with the set of stochastic 
differential equations in the rotating frame, associated with the truncated Fokker-Planck equation for three modes
\begin{eqnarray}
&& \frac{d \alpha_{0}}{dt} = -\gamma_{0} \alpha_{0} + E - \chi \alpha_{1} \alpha_{2} + \sqrt{\gamma_{0}} \xi_{0}(t)
\nonumber \\ 
&& \frac{d \alpha_{1}}{dt} = -\gamma_{1} \alpha_{1} + \chi \alpha_{0} \alpha_{2}^{*} + \sqrt{\gamma_{1}} \xi_{1}(t) 
\nonumber \\
&& \frac{d \alpha_{2}}{dt} = -\gamma_{2} \alpha_{2} + \chi \alpha_{0} \alpha_{1}^{*} + \sqrt{\gamma_{2}} \xi_{2}(t) 
\end{eqnarray}
The adiabatic elimination of the pump mode requires a special 
care. 
We need to keep the fluctuations in the steady state 
beacause they provide important additional coupling
between the down-converted modes.
The stationary solution of the pumped mode will be taken as 
\begin{equation}
\alpha_{0}= 
\frac{1}{\gamma_{0}}
\left( E - \chi \alpha_{1} \alpha_{2} +
           \sqrt{\gamma_{0}} \xi_{0}(t)\right)
\label{8B}
\end{equation}
We now substitute this expression into the other two amplitude 
mode equations to have the effective dynamical nonlinear 
equations,
\begin{eqnarray}
\frac{d \alpha_{1}}{dt} &=& -\gamma_{1} \alpha_{0} + 
\chi \alpha_{2}^{*}\frac{1}{\gamma_{0}}\left(E -\chi \alpha_{1} \alpha_{2} +\sqrt{\gamma_{0}} \xi_{0} \right) + 
\sqrt{\gamma_{1}} \xi_{1} \nonumber \\
\frac{d \alpha_{2}}{dt} &=& -\gamma_{2} \alpha_{0} + \chi \alpha_{1}^{*} 
\frac{1}{\gamma_{0}}\left(E -\chi \alpha_{1} \alpha_{2} +\sqrt{\gamma_{0}} \xi_{0} \right) + \sqrt{\gamma_{2}} \xi_{2}
\nonumber\\
\label{eq32}
\end{eqnarray}
In the vacuum state $\langle \alpha_{i}\alpha^{*}_{i} \rangle = 1/2$, and 
$\langle \xi_{i}(t)^{*} \xi_{j}(t^{\prime}) \rangle = \delta_{ij}\delta(t-t^{\prime})$. 
It is clear from the above equations that the pump noise 
enhances the coupling between the amplitudes (quadratures) of 
the down-converted modes.

In terms of real-quadrature variables,
\begin{eqnarray}
x_{1} =  \alpha_{1} + \alpha_{1}^{*} \,\,\,\,\,\,\,\,\,\, y_{1} = -i\left( \alpha_{1} + 
\alpha_{1}^{*} \right )
\nonumber \\
x_{2} =  \alpha_{2} + \alpha_{2}^{*} \,\,\,\,\,\,\,\,\,\, y_{2} = -i\left( \alpha_{2} + 
\alpha_{2}^{*} \right )
\end{eqnarray}
%
the stochastic equations (\ref{eq32}) can be written in the form
\begin{equation}
\frac{d \bold{X}}{dt} = \bold{A} + \bold{B}\, \mathbf{\eta}(t)
\end{equation}
where $\bold{X}$ is a vector that has the four quadratures as components, and the drift vector is defined as 
\begin{equation}
\bold{A}=
\gamma \left[
\begin{array}{c}
-x_{1} + \mu x_{2} - \frac{g^{2}}{2} x_{1} \left(x_{2}^{2} +y_{2}^{2}\right)  \\
\noalign{\medskip}
-y_{1} - \mu y_{2} - \frac{g^{2}}{2} y_{1} \left(x_{2}^{2} +y_{2}^{2}\right)  \\
\noalign{\medskip}
-x_{2} + \mu x_{1} - \frac{g^{2}}{2} x_{2} \left(x_{1}^{2} +y_{1}^{2}\right)  \\
-y_{2} - \mu y_{1} - \frac{g^{2}}{2} y_{2} \left(x_{1}^{2} +y_{1}^{2}\right)  \\
\end{array}
\right]
\end{equation}
where we have defined the adimensional pump parameter 
$\mu = \frac{\chi E}{\gamma \gamma_{0}}$, and the nonlinear 
coupling $g = \frac{\chi}{\sqrt{2 \gamma \gamma_{0}}}$. 
$\bold{B}$ is a $4\times 6$ matrix defined as
\begin{equation}
\bold{B} = 
\sqrt{2 \gamma} \left(
\begin{array}{cccccc}
1 &0 &0 &0 & \frac{g}{\sqrt{2}}x_{2} & \frac{g}{\sqrt{2}}y_{2} \\
0 &1 &0 &0 & -\frac{g}{\sqrt{2}}y_{2} & \frac{g}{\sqrt{2}}x_{2} \\
0 &0 &1 &0 & \frac{g}{\sqrt{2}}x_{1} & \frac{g}{\sqrt{2}}y_{1} \\
0 &0 &0 &1 & -\frac{g}{\sqrt{2}}y_{1} & \frac{g}{\sqrt{2}}x_{1} \\
\end{array}
\right)
\end{equation}
and $ \bold{\eta}(t) $ is a vector with six uncorrelated real gaussian noise components.

Associated with this set of stochastic equation there is a Fokker-Planck equation that can be written as
\begin{equation}
\frac{\partial P(\bold{X})}{\partial t} = \left[ -\frac{\partial}{\partial \bold{X}_{i}} \bold{A}_{i} + 
\frac{1}{2}\frac{\partial}{\partial \bold{X}_{i}} \frac{\partial}{\partial \bold{X}_{j}} \bold{D}_{ij}\right]
P(\bold{X})
\end{equation}
where the diffusion matrix is defined as $\bold{D}= \bold {B} \bold {B}^{T}$ and has the following expression
\begin{equation}
\bold{D} = 2 \gamma \left(
\begin{array}{cccc}
a & 0 & c & d \\
0 & a & -d & c \\
c & -d & b & 0\\
d & c & 0 & b\\
\end{array}
\right)
\end{equation}
Its inverse reads
\begin{equation}
\bold{D^{-1}} = \frac{2}{\gamma \left( ab-c^{2}-d^{2} \right)} \left(
\begin{array}{cccc}
b & 0 & -c & -d \\
0 & b & d & -c \\
-c & d & a & 0\\
-d & -c & 0 & a\\
\end{array}
\right)
\end{equation}

where we have defined
\begin{eqnarray}
a &=& 1 + \frac{g^{2}}{2}\left(x_{2}^{2} + y_{2}^{2}\right) \nonumber \\
b &=& 1 + \frac{g^{2}}{2}\left(x_{1}^{2} + y_{1}^{2}\right) \nonumber \\
c &=& \frac{g^{2}}{2}\left(x_{1}x_{2} + y_{1}y_{2}\right)   \nonumber \\
d &=& \frac{g^{2}}{2}\left(x_{1}y_{2} - y_{1}x_{2}\right) 
\end{eqnarray}
and $ab-c^{2}-d^{2} = 1 + \frac{g^{2}}{2}\left(x_{1}^{2} + y_{1}^{2} + x_{2}^{2} + y_{2}^{2} \right)$

The Fokker-Planck equation admits a potential steady state solution in the form 
$P(\bold{X})= N \exp{-\int \sum_{i} \bold{Z}_{i} \bold{X}_{i}}$, where $\bold{Z}_{i}$ given by
\begin{equation}
\bold{Z}_{i} = \sum_{k} \bold{D}_{ik}^{-1} \left[2 \bold{A}_{k} - \sum_{j} \frac{\partial}{\partial \bold{X}_{j}}
\bold{D}_{ij} \right]
\end{equation}
if the potential condition, $\, \partial_{i} \bold{Z}_{j} = \partial_{j} \bold{Z}_{i}\,$, is satisfied.

In the present case the $\bold{Z}$ vector is written as
\begin{eqnarray}
Z_{1} &=& \frac{1}{1+\frac{g^{2}}{2}(x_{1}^{2}+y_{1}^{2}+x_{2}^{2}+y_{2}^{2})} \times\nonumber\\
& &\left[ -(1+g^{2}) x_{1} + \mu x_{2} - \frac{g^{2}}{2} x_{1} \left(x_{1}^{2}+y_{1}^{2} \right)  \right] \nonumber \\
Z_{2} &=& \frac{1}{1+\frac{g^{2}}{2}(x_{1}^{2}+y_{1}^{2}+x_{2}^{2}+y_{2}^{2})} \times\nonumber\\ 
& &\left[ -(1+g^{2}) y_{1} + \mu y_{2} - \frac{g^{2}}{2} y_{1} \left(x_{1}^{2}+y_{1}^{2} \right)  \right] \nonumber \\
Z_{3} &=& \frac{1}{1+\frac{g^{2}}{2}(x_{1}^{2}+y_{1}^{2}+x_{2}^{2}+y_{2}^{2})} \times\nonumber\\ 
& &\left[ -(1+g^{2}) x_{2} + \mu x_{1} - \frac{g^{2}}{2} x_{2} \left(x_{2}^{2}+y_{2}^{2} \right)  \right] \nonumber \\
Z_{4} &=& \frac{1}{1+\frac{g^{2}}{2}(x_{1}^{2}+y_{1}^{2}+x_{2}^{2}+y_{2}^{2})} \times\nonumber\\ 
& &\left[ -(1+g^{2}) y_{2} + \mu y_{1} - \frac{g^{2}}{2} y_{2} \left(x_{2}^{2}+y_{2}^{2} \right)  \right]
\end{eqnarray} 

Unfortunately, a potential solution cannot be achieved with the present form of matrix $\bold{D}$. 
Of course, a steady state solution of the Fokker-Planck equation still exists, since the detailed 
balance is established, but the solution will not be in a potential form. 

For a sufficiently peaked distribution, we may approximate the diffusion matrix $\bold{D}$ by 
its average. 
The off diagonal terms have zero mean value due to phase space symmetry. Indeed, it is easy to 
verify that the dynamical equations remain unchanged under the transformation 
$\alpha_{1}^{\prime} = \alpha_{1} \exp (-i \theta)$ and $\alpha_{2}^{\prime} = \alpha_{2} \exp (i \theta)$. 
which corresponds to a definition of rotated quadratures $x_1^{\prime},y_1^{\prime},x_2^{\prime},y_2^{\prime}$. 
This symmetry of the dynamical equations imply in 
$W(x_1,y_1,x_2,y_2)=W(x_1^{\prime},y_1^{\prime},x_2^{\prime},y_2^{\prime})$
For a particular choice of $\theta = \pi/2$, this corresponds $x_1^{\prime}=y_1$, $y_1^{\prime}=-x_1$, 
$x_2^{\prime}=-y_2$, and $y_2^{\prime}=x_2$ so that $W(x_1,y_1,x_2,y_2)=W(y_1,-x_1,-y_2,x_2)$. This 
parity property of the Wigner function forces the off diagonal terms $c$ and $d$ to be zero. 
Note that the diagonal terms of the diffusion matrix are related to the individual intensities of 
the down-converted fields and can be replaced by the mean values calculated from the deterministic 
classical equations. This substantially simplifies the expression for vector $\bold{Z}$

\begin{eqnarray}
Z_{1} &=& \frac{1}{s(\mu)} \left[ - x_{1} + \mu x_{2} 
- \frac{g^{2}}{2} x_{1} \left(x_{2}^{2}+y_{2}^{2} \right)  \right] \nonumber \\
Z_{2} &=& \frac{1}{s(\mu)} \left[ - y_{1} + \mu y_{2} 
- \frac{g^{2}}{2} y_{1} \left(x_{2}^{2}+y_{2}^{2} \right)  \right] \nonumber \\
Z_{3} &=& \frac{1}{s(\mu)} \left[ - x_{2} + \mu x_{1} 
- \frac{g^{2}}{2} x_{2} \left(x_{1}^{2}+y_{1}^{2} \right)  \right] \nonumber \\
Z_{4} &=& \frac{1}{s(\mu)} \left[ - y_{2} + \mu y_{1} 
- \frac{g^{2}}{2} y_{2} \left(x_{1}^{2}+y_{1}^{2} \right)  \right]
\end{eqnarray} 
where 
$s(\mu) = 1 + g^{2}\, \langle x_{i}^{2} + y_{i}^{2} \rangle /2$
is given by Eq.~(\ref{eq15}).

This vector represents a conservative field that generates the approximate potential solution of the 
effective Fokker-Planck equation.

\end{document}